\documentclass[onecollarge]{svjour2}
\usepackage{graphicx}
\usepackage{cite}
\usepackage{color}
\usepackage{amsfonts}
\usepackage{soul}
\usepackage{cancel}

\begin{document}

\title{A Gallavotti-Cohen-Evans-Morriss like symmetry for a class of Markov jump processes}

\author{Andre Cardoso Barato \and Rapha\"el Chetrite \and Haye Hinrichsen \and David Mukamel}

\institute{
A. C. Barato \at
The Abdus Salam International Centre for Theoretical Physics, Trieste 34014, Italy\\
\email{acardoso@ictp.it}\\
\and
R. Chetrite 
\at  Laboratoire J. A. Dieudonn\'e, UMR CNRS 6621, Universit\'e de Nice Sophia-Antipolis, Parc Valrose, 06108 Nice Cedex 02, France \\ 
\email{Raphael.Chetrite@unice.fr}\\
\and
H. Hinrichsen 
\at Fakult{\"a}t f{\"u}r Physik und Astronomie, Universit{\"a}t W{\"u}rzburg, Am Hubland, 97074 W{\"u}rzburg, Germany\\ 
\email{hinrichsen@physik.uni-wuerzburg.de}\\
\and
D. Mukamel \at Department of Physics of Complex Systems, Weizmann Institute of Science, Rehovot 76100, Israel\\ 
\email{david.mukamel@weizmann.ac.il}\\
}

\date{Received: date / Accepted: date}

\maketitle

\begin{abstract}
We investigate a new symmetry of the large deviation function of certain time-integrated currents in non-equilibrium systems. The symmetry is similar to the well-known Gallavotti-Cohen-Evans-Morriss-symmetry for the entropy production, but it concerns a different functional of the stochatic trajectory. The symmetry can be found in a restricted class of Markov jump processes, where the network of microscopic transitions has a particular structure and the transition rates satisfy certain constraints. We provide three physical examples, where time-integrated observables display such a symmetry. Moreover, we argue that the origin of the symmetry can be traced back to time-reversal if stochastic trajectories are grouped appropriately.
 
 
\end{abstract}


\def\conf{c}					
\def\rate{w}					
\def\headline#1{\paragraph{\bf #1\\[2mm]}}	
\def\multiindex#1{\mathbf{#1}}			

\newpage
\parskip 2mm 
\pagestyle{plain}

\section{Introduction}

In Nature most processes are not in thermodynamic equilibrium. For example, whenever a system is exposed to a flux of matter or energy in the stationary state, then it is generally not possible to describe it with standard methods of equilibrium statistical mechanics. For systems out of equilibrium the probability distribution of configurations is normally not known and a theory which allows one to calculate macroscopic quantities is not available. Therefore, general results valid for nonequilibrium processes are of great theoretical interest. This applies in particular to fluctuation relations \cite{evans93,evans94,gallavotti95,kurchan98,lebowitz99,maes99,Jia1,andrieux07,harris07,kurchan07}, which are fairly general statements valid for any system out of equilibrium.   

A large variety of nonequilibrium systems can be modelled as continuous-time Markov jump processes, meaning that the system jumps spontaneously from one classical configuration to the other with certain rates. If these rates obey detailed balance, the system relaxes into a equilibrium state. However, if detailed balance is broken, the system will relax into a nonequilibrium stationary state where the probability current between microstates is non-zero \cite{zia07}. Maintaining these non-vanishing probability currents requires an external drive which continuously produces entropy in the environment. Remarkably, this entropy production can be quantified without knowing the explicit structure of the environment~\cite{schnakenberg76}. It turns out that the average entropy production is zero if and only if detailed balance is fulfilled, meaning that entropy production can be used as an indicator of nonequilibrium. 

Since the entropy production depends on the specific sequence of microscopic transitions (the so-called stochastic path), it is a fluctuating quantity with a certain probability distribution~\cite{seifert05}. A fluctuation relation is an equation that restricts the functional form of this distribution. There are two different kinds of fluctuation relations, namely, finite time fluctuation relations and infinite time fluctuation relations (see \cite{harris07}). Finite time fluctuation relations describe the distribution of the total entropy (system + environment) and hold exactly for any time interval~\cite{seifert05}. Relations of this kind include the Jarzynski equality \cite{jarzynski97} and Crooks relation \cite{croocks99}. On the other hand, infinite time fluctuation relations are asymptotically valid for the entropy produced in the environment (which is also known as action functional) \cite{lebowitz99}. Here we will deal with this infinite time relation, and we refer to it as the fluctuation  theorem or the Gallavotti-Cohen-Evans-Morriss (GCEM) symmetry.  

Large deviation theory \cite{ellis85,touchette09,hollander,oono} is the appropriate mathematical framework to investigate the fluctuation theorem. Using these methods the GCEM symmetry can be recast as a symmetry of the large deviation function for the probability distribution of the entropy. This symmetry does not yet allow us to calculate macroscopic observables, but at present it is the most general result for systems out of equilibrium. 

Even though the fluctuation theorem is very general, one might argue that it is also very specific in the sense that it is valid only for one particular functional of the stochastic path, namely, the entropy produced in the environment. It is therefore interesting to find out if there are other physically relevant functionals with a similar symmetry. As a first step in this direction, we recently demonstrated that the height of an interface in a certain growth model defines a physically relevant time-integrated current different from the entropy with a symmetric large deviation function~ \cite{barato10}. Interestingly, this symmetry can only be observed for a particular system size of the model because only then the network of microscopic transitions acquires a particular form. 

The objective of the present paper is to generalize the new symmetry found in \cite{barato10}. We prove that for a class of jump processes with a particular network of states we can find time-integrated currents different from entropy displaying a symmetric large deviation function. This symmetry is similar to the GCEM symmetry because it restricts the form of a large deviation function related to a time-integrated current. However, it is important to note that our symmetry is different from the GCEM symmetry because it refers to a different time-integrated current and has a slightly different physical origin.

It is known that the GCEM symmetry is a direct consequence of the fact that the entropy is given by the weight of a stochastic path divided by the weight of the time-reversed path. In this sense, the origin of the GCEM symmetry is related to time-reversal. An interesting question, which we address here for a particular case, would be to explain the origin of the symmetry. We show that it is also associated with time-reversal, but in a more hidden way: it comes to light only when we perform an appropriate grouping of stochastic trajectories.  

The organization of the paper is as follows. In Sec. 2 we define time-integrated currents and briefly review the fluctuation theorem. In Sec. 3 we prove the new symmetry relation. Some physical examples where our symmetry appears in physically meaningful time-integrated currents are presented in Sec.~4. Before concluding we discuss the origin of the symmetry for a simple four states system in Sec. 5.

\section{Time-integrated currents and the fluctuation theorem}

Continuous-time Markov jump processes are defined by a space of microscopic configurations $\conf\in\Omega$ in which the system evolves by spontaneous transitions $\conf \to \conf'$ at rate $\rate_{\conf \to \conf'}$. The probability $P(\conf,t)$ of finding such a system at time $t$ in the configuration $\conf$ evolves according to the master equation
\begin{equation}
\frac{d}{dt}P(\conf,t)= \sum_{\conf'\neq \conf}\Big(P(\conf',t)\rate_{\conf'\to \conf}-P(\conf,t)\rate_{\conf\to \conf'}\Big).
\label{master}
\end{equation}
For simplicity we assume that stationary probability distribution exists  $P(\conf)=\lim_{t\to\infty}P(\conf,t)$. If this distribution obeys the condition of detailed balance  $P(\conf)\rate_{\conf\to \conf'}= P(\conf')\rate_{\conf'\to \conf}$ the stationary state is an equilibrium state, otherwise the system is out of equilibrium.

A stochastic trajectory during the time interval $\left[t_0,t_f\right]$ is a sequence of $M$ jumps 
\begin{equation}
\overrightarrow{C}_{M,t}: \, \conf(t_0) \to \conf(t_1) \to \conf(t_2) \to \ldots \to \conf(t_M)
\end{equation}
taking place at times $t_1,t_2,\ldots,t_M \in \left[t_0,t_f\right]$. Note that, the length of the time interval $T= t_f-t_0$ is given while the number of jumps $M$ is a random variable that may assume different values for different trajectories. In what follows we assume that all microscopic transitions are such that, i.e. $\rate_{\conf\to \conf'}\neq 0 \, \Leftrightarrow \, \rate_{\conf'\to \conf}\neq 0$, meaning that any stochastic path can be reversed.

\headline{Time-integrated currents}
A time-integrated current is a functional of the stochastic trajectory 
\begin{equation}
J[\overrightarrow{C}_{M,t}]=\sum_{i=0}^{M-1}\theta _{\conf(t_i)\to \conf(t_{i+1})}\,,
\label{defcurrent}
\end{equation}
which changes its value by $\theta_{\conf\to\conf'}$ whenever a jump from $\conf\to\conf'$ occurs. The increments are assumed to be antisymmetric, i.e. $\theta _{\conf\to \conf^{\prime }}=-\theta _{\conf^{\prime }\to \conf}$. Using the master equation the expectation value is
\begin{equation}
\left\langle J\right\rangle =\int_{t_0}^{t_f}dt
\sum_{\conf,\conf^{\prime }}\theta_{\conf\rightarrow \conf^{\prime }}P(\conf,t)w _{\conf\rightarrow \conf^{\prime }}\,.
\end{equation}
In the stationary state this expression reduces to $\left\langle J\right\rangle = T
\sum_{\conf,\conf^{\prime }}\theta_{\conf\rightarrow \conf^{\prime }}P(\conf)w _{\conf\rightarrow \conf^{\prime }}$, i.e. the current increases on average linearly with $T$. Since the system relaxes towards a stationary state, in the limit of $T\to\infty$ the quotient $J/T$ tend to a constant, which is
\begin{equation}
\lim_{T\to \infty}\frac{ J }{T}\,\rightarrow  \sum_{\conf,\conf^{\prime }}\theta _{\conf\rightarrow
\conf^{\prime }}P(\conf)w_{\conf\rightarrow \conf^{\prime }}.
\end{equation}
More specifically, one expects that the corresponding probability distribution $P\left(\frac{J}{T}=x\right)$ becomes more and more peaked around this value as $T\to \infty$. Assuming that the large deviation principle holds, the large deviation function of this probability distribution is defined by \cite{ellis85,touchette09,hollander},
\begin{equation}
\lim_{T\to\infty}P\left(\frac{J}{T}=x\right) = \exp [-TI(x)].
\label{rate}
\end{equation}
The function $I(x)$ measures the rate at which the current deviates from its average value.

\headline{Entropy production and fluctuation theorem}
A very prominent time-integrated current is the entropy. If we have a jump process describing a physical system in contact with external reservoirs, this quantity describes the amount of entropy which is generated by the external driving in a fictitious external environment~\cite{schnakenberg76,seifert05,hinrichsen11}. More specifically, each transition $\conf\to\conf'$ changes the entropy in the environment by $\ln\frac{\rate_{\conf\to\conf'}}{\rate_{\conf'\to\conf}}$. Therefore, the accumulated entropy production along a stochastic path $\overrightarrow{C}_{M,t}$ is a time-integrated current of the form
\begin{equation}
\label{entropy}
J_s[\overrightarrow{C}_{M,t}] =\sum_{i=0}^{M-1} \ln \frac{\rate_{\conf(t_i)\to\conf(t_{i+1})}}{\rate_{\conf(t_{i+1})\to\conf(t_i)}}\,.
\end{equation}
The fluctuation theorem reads ~\cite{lebowitz99}
\begin{equation}
I_s(x)-I_s(-x)=-x,
\label{GCEMforlarge}
\end{equation}
where $I_s(x)$ is the large deviation function associated with $\lim_{T\to\infty}P(\frac{J_s}{T}=x)$. This property of the entropic current is also referred to as the GCEM symmetry.

\headline{Determining the large deviation function}
The master equation (\ref{master}) can be rewritten in the form
\begin{equation}
\frac{d}{dt}P(\conf,t)= -\sum_{\conf'}\hat\mathcal{L}_{\conf\conf'}P(\conf',t),
\end{equation}
where $\hat\mathcal{L}$ is the Markov generator with elements
\begin{equation}	
\hat{\mathcal{L}}_{\conf\conf'}=\left\{
\begin{array}{ll} 
- w_{\conf'\to \conf} & \quad \textrm{ if } \conf\neq \conf'\\
 \lambda(\conf) & \quad \textrm{ if }  \conf=\conf' 
\end{array}\right.\,.
\end{equation}
Here 
\begin{equation}
\lambda(\conf)=\sum_{\conf'\neq \conf} \rate_{\conf\to \conf'}
\label{escaperate}
\end{equation}
denotes the \textit{escape rate} from configuration $\conf$. For each time-integrated current of the form (\ref{defcurrent}), one can now define a modified generator  $\hat\mathcal{L}(z)$ by 
\begin{equation}	
\hat{\mathcal{L}}(z)_{\conf\conf'}=\left\{\begin{array}{ll} 
 -w_{\conf'\to \conf}\exp(-z \,\theta_{\conf'\to \conf}) & \quad \textrm{if } \conf\neq \conf'\\
 \lambda(\conf) & \quad \textrm{if } \conf=\conf'
\end{array}\right.\,.
\label{modgenerator}
\end{equation}
The scaled cumulant generating function $\hat{I}(z)$, of the respective time-integrated current $J$, is defined by
\begin{equation}
\lim_{T\to\infty}\left\langle \exp (-zJ)\right\rangle = \exp (- T\hat{I}(z)).
\end{equation}
It can be shown that $\hat{I}(z)$ is given by the minimum eigenvalue of the modified generator (\ref{modgenerator})  \cite{lebowitz99}. Moreover, the Gr\"atner-Ellis theorem~\cite{ellis85,touchette09,hollander} states that $I(x)$ is given by the Legendre-Fenchel transform of $\hat{I}(z)$, i.e.,
\begin{equation}
I(x)=\textrm{max}_z\left(\hat{I}(z)-xz\right)\,,
\end{equation}
with $z$ real. Note that $\hat{I}(0)=0$ because in this case $\mathcal{L}(z)$ reduces to the Markov generator $\mathcal{L}$ with the minimum eigenvalue $0$. The GCEM symmetry (\ref{GCEMforlarge}) in terms of the scaled cumulant generating function of the entropy $\hat{I}_s(z)$ is
\begin{equation}
\hat{I}_s(z)=\hat{I}_s(1-z).
\label{GCEMnonoscaled}
\end{equation}   
The advantage of dealing with the scaled cumulant generating function is that it is easier to calculate in several situations. In the present case it corresponds to determine the minimum eigenvalue of the Perron-Frobenius matrix (\ref{modgenerator}). Note that, the curve $\hat{I}_s(z)$ has a convex shape, vanishes at $z=0$ and $z=1$,  and reaches its minimum at $z=1/2$.

\headline{Non-entropic time-integrated currents}
Is it possible to find other time-integrated currents with a GCEM-like symmetry? Previous works (see e.g.  \cite{harris07, bodineau07}) have shown that such currents do exist. However, these examples are unsatisfactory in so far as the proposed currents differ from the entropy only initially while they become proportional to the entropy in the long time limit. If we call such a current $J_r$, this means that $\hat{I}_r(z)=\hat{I}_r(E-z)$. Moreover, since the current $J_r$ becomes proportional to entropy in the long time limit, $\hat{I}_r(zE)=\hat{I}_s(z)$, i.e., the rescaled scaled cumulant generating functions of $J_r$ and  $J_s$ have the same functional form.

The main objective of the present paper is to show that it is possible to find symmetric currents which differ from the entropy even in the limit $T\to \infty$ so that their rescaled scaled cumulant generating functions differ from $\hat{I}_s(z)$, even tough they are both symmetric and touch the horizontal axis at the same points (see below in Figs. \ref{legendremechanical}, \ref{legendreRSOS}, and \ref{legendreDavid}).

\headline{Counting the degrees of freedom}
Before proceeding let us point out that there are certain restrictions that reduce the degrees of freedom in the space of time-integrated currents. As such currents are specified by an antisymmetric matrix $\theta_{\conf\conf'}=-\theta_{\conf'\conf}$ there are in principle $N(N-1)/2$ degrees of freedom, where $N$ is the number of states. However, not all of them are independent. To see this let us consider the elementary time-integrated currents
\begin{equation}
J_{\conf \to \conf'}[\overrightarrow{C}_{M,t}] = \sum_{i=0}^{M-1} \left( \delta_{\conf,\conf(t_i)}\delta_{\conf',\conf(t_{i+1})}-\delta_{\conf',\conf(t_i)}\delta_{\conf,\conf(t_{i+1})} \right)
\label{defelementary}
\end{equation}
from which all other currents can be constructed by linear combination. The current $J_{\conf \to \conf'}[\overrightarrow{C}_{M,t}]$ is simply the number of transitions $\conf\to\conf'$ minus the number of reverse transitions $\conf'\to\conf$ along the stochastic path $\overrightarrow{C}_{M,t}$. The sum over all destinations
\begin{equation}
N_\conf = \sum_{\conf'} J_{\conf \to \conf'}
\end{equation}
is just the number how often the system reaches the configuration $c$ minus the number how often this configuration is left, hence $N_\conf$ can only take the values 0 and $\pm 1$. This implies that 
\begin{equation}
\sum_{\conf'}\frac{J_{\conf \to\conf'}}{T}\to 0.
\label{restcurr}
\end{equation} 
in the limit $T \to \infty$, meaning that these particular linear combinations of elementary currents do not contribute to the large-deviation function. This reduces the degrees of freedom by the number of independent relations (\ref{restcurr}), which is maximally $N$.

\section{A symmetric time-integrated current different from entropy}
\label{sec3}
\begin{figure}
\begin{center}	
\includegraphics[width=133mm]{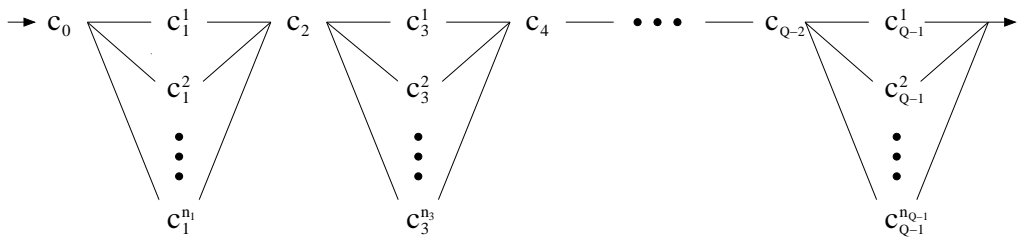}
\caption{Artificial network of configurations $\conf_k^i$ with periodic boundary conditions. The process can jump between configurations that are connected by a line. } 
\label{networkgeneral}
\end{center}
\end{figure}
In this section we prove that for jump processes with a particular network of states and suitably chosen transition rates one can define a current with a symmetric large deviation function which differs from the one for the entropy. The structure of this network is shown in Fig.~\ref{networkgeneral}. It consist of configurations $c_k^i$ organized in columns labeled by a lower index $k=0,\dots,Q-1$, each of them including $n_k$ different configurations labeled by an upper index $i=1\ldots n_k$. Spontaneous jumps are allowed only between configurations in neighboring columns with periodic boundary conditions, as indicated by straight lines in the figure. Moreover, we assume that the number of columns $Q$ is even and that even columns carry only a single configuration, i.e. $n_0=n_2=n_4=\ldots=1$. This forces the system to go through periodically arranged bottlenecks of single configurations.

On this network of configurations we consider the current
\begin{equation}
\label{ourcurrent}
J_r \;=\; \sum_{k=0}^{Q-1} \theta_k \sum_{i_k=1}^{n_k} \sum_{i_{k+1}=1}^{n_{k+1}} J_{\conf_k^{i_k} \to \conf_{k+1}^{i_{k+1}}}\,,
\end{equation}
where $\theta_k$ are numbers and $J_{\conf\to \conf'}$ are the elementary currents defined in Eq.~(\ref{defelementary}). This current increases by $\theta_k$ if the process process jumps from column $k$ to $k+1$ and decreases by the same amount if the system jumps in the opposite direction, no matter which of the configurations within a column is selected.

\headline{Counting cycles}
Assume that the process starts at a particular configuration, say $\conf_0$. Depending on the yet unspecified transition rates, the process will perform a random walk from column to column. Whenever it returns to its starting point $\conf_0$, it is easy to see that the current defined above will take the value $J_r=m\Theta$, where $\Theta=\sum_{k=0}^{Q-1}\theta_k$ and $m\in\mathbb{Z}$. The number $m$ tells us how often the system completed a cycle through all columns $0\to 1\to 2 \to\ldots\to Q-1\to 0$. Therefore, if the rates are chosen in such a way that the random walk through the columns is biased to the right, $m$ will be on average positive. With this picture in mind it is intuitively clear that the expectation value of $m$ in the long-time limit is related to the average current through each of the bottlenecks. 

To prove this intuitive argument, we apply the restriction~(\ref{restcurr}) to each of the configurations in the network. For single but multiply connected configurations at even columns this restriction tells us that the incoming current is equal to the outgoing current in the long time limit, i.e. for even $k$ we have
\begin{equation}
\sum_{i=1}^{n_{k-1}}\frac{J_{\conf_{k-1}^{i}\to \conf_k}}{T}=\sum_{i=1}^{n_{k+1}}\frac{J_{\conf_{k}\to \conf_{k+1}^{i}}}{T}\,.
\label{rest2}
\end{equation}
For the simply connected configurations in columns with odd $k$, we have instead
\begin{equation}
\frac{J_{\conf_k^{i}\to \conf_{k+1}}}{T}= \frac{J_{\conf_{k-1}\to \conf_k^{i}}}{T}\,,
\label{rest1}
\end{equation}
where $i=1,\ldots,n_k$. From the above relations it is easy to show that the current in the large deviation regime is given by
\begin{equation}
\frac{J_r}{T}= \Theta\sum_{i=1}^{n_1}\frac{J_{\conf_0\to \conf_1^{i}}}{T}\,.
\label{1cld}
\end{equation}    
This result proves that the large deviation properties of the current do not depend individually on the contributions $\theta_k$ but only on their sum $\Theta=\sum_k\theta_k$. This means that all currents of the form~(\ref{ourcurrent}) are proportional in the long-time limit and therefore characterized (up to rescaling) by the same large deviation function.

\headline{Structure of the characteristic polynomial}
We now prove that for suitably chosen transition rates the scaled cumulant generating function of $J_r$ exhibits the symmetry (\ref{GCEMnonoscaled}). Following  Ref.~\cite{andrieux07} we first consider the characteristic polynomial
\begin{equation}
\label{charpol1}
P(z,x)\;=\;\det\bigl(x I-\hat\mathcal{L}(z)\bigr) \;=\; \sum_\pi \mbox{sgn}(\pi) \prod_c \Bigl(x\delta_{c,\pi_c}-\hat\mathcal{L}(z)_{c,\pi_c} \Bigr)\,
\end{equation}
where the sum runs over all permutations $\pi$ of the available configurations. Here $\hat\mathcal{L}(z)$ denotes the modified generator (\ref{modgenerator}) with the matrix elements
\begin{equation}
\hat\mathcal{L}(z)_{\conf',\conf} \;=\; -\rate_{\conf\to\conf'}\exp(-z\,\theta_{\conf\to\conf'})+\delta_{\conf,\conf'}\lambda(\conf)
\end{equation}
\begin{figure}
\begin{center}	
\includegraphics[width=130mm]{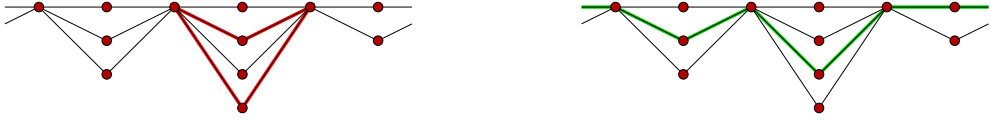}
\caption{Local transition cycle (left) and complete transition cycle through all columns (right).} 
\label{cycles}
\end{center}
\end{figure}
where $\lambda(\conf)$ is again the escape rate~(\ref{escaperate}). Since the off-diagonal entries represent the possible microscopic transitions only those permutations will contribute to the determinant which correspond to a set of non-intersecting transition cycles. In the network of configurations shown in Fig.~\ref{networkgeneral} there are three types of closed transition cycles, namely local cycles which do not change the current $J_r$, and cycles extending over the whole system in positive or negative direction, changing the current by $\pm \Theta$  (see Fig.~\ref{cycles}). In the determinant this means that permutations corresponding to local cycles are $z$-independent since the exponential factors drop out. Conversely, complete cycles extending over the whole system contribute to the sum with terms which are proportional to $e^{\pm z\Theta}$, respectively. We can therefore split the characteristic polynomial (\ref{charpol1}) into three parts
\begin{equation}
\label{pol}
P(z,x)\;=\;f(x) e^{-z \Theta} + \overline f(x) e^{z \Theta} + g(x)\,.
\end{equation}
Labelling each complete cycle $\conf_0\to\conf_1^{i_1}\to\conf_2\to\conf_3^{i_3}\ldots\to\conf_{Q-1}^{i_{Q-1}}\to\conf_0$ by a multiindex $\multiindex{i}:=(i_1,i_3,\ldots,i_{Q-1})$
the first two functions can be expressed as
\begin{equation}
\label{fandfbar}
f(x) \;=\; \sum_{\multiindex i} T_{\multiindex i} R_{\multiindex i}(x)\,,\qquad
\overline f(x) \;=\; \sum_{\multiindex i} \overline T_{\multiindex i} R_{\multiindex i}(x)
\end{equation}
where $T_{\multiindex i}$ and $\overline T_{\multiindex i}$ are the products of all rates along the cycle in forward and backward direction, respectively. Explicitly, 
\begin{equation}
T_{\multiindex i}= \rate_{\conf_0\to \conf_{1}^{i_{1}}}\rate_{\conf_{1}^{i_{1}}\to \conf_2}\ldots\rate_{\conf_{Q-2}\to \conf_{Q-1}^{i_{Q-1}}}\rate_{\conf_{Q-1}^{i_{Q-1}}\to \conf_0}\,, \qquad
\overline T_{\multiindex i}=\rate_{\conf_0\to\conf_{Q-1}^{i_{Q-1}}}\rate_{\conf_{Q-1}^{i_{Q-1}}\to\conf_{Q-2}}\ldots\rate_{\conf_2\to\conf_{1}^{i_{1}}}\rate_{\conf_{1}^{i_{1}}\to \conf_0}.
\end{equation}
Moreover, $R$ comprises all diagonal entries coming from the configurations that are not involved in the cycle:
\begin{equation}
\label{Req}
R_{\multiindex i}(x) = \prod_{k=1}^{Q/2} \,\prod_{j_{2k-1}=1, j_{2k-1}\neq i_{2k-1}}^{n_{2k-1}} \Bigl( -\lambda(\conf_{2k-1}^{j_{2k-1}})+x \Bigr)
\end{equation}
There is also a complicated expression for $g(x)$ but, as we will see below, this function is not needed for finding the symmetry.

\headline{Symmetry condition and derivation of the constraints}
Clearly, a symmetry of the characteristic polynomial $P(z,x)=P(E-z,x)$ is a sufficient condition for the symmetry $\hat{I}_r(z)=\hat{I}_r(E-z)$ of the minimal eigenvalue. Because of Eq.~(\ref{pol}) this means that the condition
\begin{equation}
f(x)=\exp(E\Theta)\overline{f}(x)
\label{constraints}
\end{equation}
implies a GCEM symmetry of the current defined in (\ref{ourcurrent}). Since $f(x)$ and $\overline f(x)$ are polynomials in $x$ of order $N-Q$, where $N=\sum_k n_k$ is the total number of configurations, we can compare the coefficients on both sides. Equating the leading order $x^{N-Q}$ we obtain
\begin{equation}
\exp(E \Theta)=\frac{\sum_{\multiindex i}T_{\multiindex i}}{\sum_{\multiindex i}\overline T_{\multiindex i}}\,.
\label{eqsymmetricfactor}
\end{equation}
The above formula says that $\exp(E \Theta)$ is given by the sum of the product of the transition rates of each possible forward cycle that  goes trough all the columns (increasing the current by $\Theta$) divided by the sum of the product of the transition rates of each possible backward cycle. 

The comparison of the $N-Q$ remaining orders yields a set of constraints for the transition rates. To this end we rewrite Eq.~(\ref{Req}) as a power series
\begin{equation}
\label{ReqPowerSeries}
R_{\multiindex i}(x) = \, \sum_{n=0}^{N-Q}(-1)^{n} x^{N-Q-n} \, \sigma_n(\Lambda_{\multiindex i})\,,
\label{defR}
\end{equation}
where $\sigma_n(Y_1,Y_2,\ldots,Y_{N-Q})=\sum_{1\le l_1<l_2<\ldots<l_n\le N-Q}Y_{l_1}Y_{l_2}\ldots Y_{l_{n}}$ is the elementary symmetric polynomial and where
\begin{equation}
\Lambda_{\multiindex i}\;=\;\Bigl\{\,\lambda(c_k^{j_k})\quad \Bigl|\quad k=1,3\ldots,Q-1; \quad j_k=1,\ldots,n_k; \quad j_k\neq i_k\Bigr\}
\label{lambdabig}
\end{equation}
is a set of $N-Q$ arguments, consisting of all escape rates which are not part of the cycle labeled by $\multiindex i$. Inserting this power series into Eq.~(\ref{fandfbar}) and comparing the coefficients in Eq.~(\ref{constraints}) one is led to $N-Q$ constraints for the transition rates of the form
\begin{equation}
\label{consym}
\frac{\sum_{\multiindex i} T_{\multiindex i}}{\sum_{\multiindex i} \overline T_{\multiindex i}} \;=\; 
\frac{\sum_{\multiindex i} T_{\multiindex i} \, \sigma_n(\Lambda_{\multiindex i})}{\sum_{\multiindex i} \overline T_{\multiindex i} \, \sigma_n(\Lambda_{\multiindex i})}
\qquad n=1,2,\ldots,N-Q 
\end{equation}

\headline{Simple solutions of the constraints}
Even though these constraints appear to be complicated, they are fulfilled trivially by setting
\begin{equation}
\lambda(\conf_k^i):=\lambda_k
\label{constressimple}
\end{equation}
for all $k=0,\ldots,Q-1$ and $i=1,\ldots,n_k$, meaning that all configurations in the same column have the same escape rate. In this case the function $R_{\multiindex i}(x)=R(x)$ does no longer depend on the specific choice of the cycle labeled by the multiindex $\multiindex i$, meaning that the polynomials $f(x)$ and $\overline{f}(x)$ reduce to
\begin{equation}
f(x)=R(x)\sum_{\multiindex i}T_{\multiindex i}\,,\qquad
\overline f(x)=R(x)\sum_{\multiindex i}\overline T_{\multiindex i}\,.
\end{equation}
Clearly these functions satisfy equation (\ref{constraints}), which implies in a symmetric characteristic polynomial and therewith a GCEM-like symmetry of the current~(\ref{ourcurrent}).

Another trivial solution that fulfills the constraints is the following. Let us take a column $k$ that has more than one state and define the quantity  
\begin{equation}
F_{k}^{i_k}= \frac{\rate_{\conf_{k-1}\to \conf_k^{i_k}}\rate_{\conf_k^{i_k}\to \conf_{k+1}}}{\rate_{\conf_{k+1}\to \conf_k^{i_k}}\rate_{\conf_k^{i_k}\to \conf_{k-1}}}\,, 
\label{contri2}
\end{equation}
where $i_k=1,2,\ldots,n_k$. If this quantity is constant in all columns with more than one site, i.e.
\begin{equation}
F_{{k}}=F_{{k}}^{i_{{k}}}\qquad\textrm{for }i_{{k}}=1,\ldots,n_{{k}},
\label{Fkconst}
\end{equation}
then the ratio
\begin{equation}
\frac{T_{\multiindex i}}{\overline T_{\multiindex i}} \;=\; \prod_{k=1}^{Q/2} F_{2k-1}
\end{equation}
is independent on the cycle labeled by the multiindex $\multiindex i$ because the reversal any complete cycle changes the product of the rates always by the same factor, hence we arrive at the symmetry condition
\begin{equation}
f(x)= \left(\prod_{k=1}^{Q/2}F_{2k-1}\right)\overline{f}(x)\,.
\end{equation}
Finally, we get a larger class of  solutions by mixing the conditions (\ref{constressimple}) and (\ref{Fkconst}), i.e., some of the $Q$ columns with more than one state have constant escape rates while the remaining columns have constant $F_{{k}}$. Using similar arguments one can again show that the ratio $f/\overline f$ is constant and, therefore, the constraints (\ref{consym}) are fulfilled. Nevertheless, we note that  not all the solutions of the constraints equations are of this type.

\headline{The relation between $J_r$ and $J_s$}
Depending on the transition rates the current $J_r$ may be proportional to entropy. If this is the case, then $J_r$ displays the GCEM symmetry. However, when $J_r$ is not proportional to entropy in the large deviation regime and still with a symmetric large deviation function, then we have a symmetry different from the GCEM symmetry. In the following we derive the condition on the transition rates such that  $J_s$ is proportional to $J_r$. 

For the network of states shown in Fig. \ref{networkgeneral} the entropy current is given by
\begin{equation}
\frac{J_s}T\;=\; \frac1T
\sum_{k=1,3,5,\ldots}^{Q-1}\,\left(
\sum_{i=1}^{n_k}J_{\conf_{k-1}\to \conf_k^{i}}\ln\frac{\rate_{\conf_{k-1}\to \conf_k^{i}}}{\rate_{\conf_k^{i}\to \conf_{k-1}}}+\sum_{i=1}^{n_k}J_{\conf_k^{i}\to \conf_{k+1}}\ln\frac{\rate_{\conf_k^{i}\to \conf_{k+1}}}{\rate_{\conf_{k+1}\to \conf_k^{i}}}
\right)\,.
\end{equation}  
Using relation (\ref{rest1}), in the long time limit the above term divided by $T$ becomes
\begin{equation}
\frac{J_s}T\;=\; \frac1T
\sum_{k=1,3,5,\ldots}^{Q-1}\,\sum_{i=1}^{n_k}J_{\conf_{k-1}\to \conf_k^{i}}\ln\frac{\rate_{\conf_{k-1}\to \conf_k^{i}}\rate_{\conf_k^{i}\to \conf_{k+1}}}{\rate_{\conf_{k+1}\to \conf_{k}^{i}}\rate_{\conf_k^{i}\to \conf_{k-1}}}\,.
\label{becentropy}
\end{equation}  
Let us first consider the second solution of the constraints (\ref{Fkconst}). In this case the quotient of rates appearing in the logarithm does not depend on $i$, therefore, using relation (\ref{rest2}) we obtain
\begin{equation}
\frac{J_s}{ T} \;=\; \frac1T \left(\sum_{k=1,3,5,\ldots}^{Q-1}\ln F_k \right) \left(\sum_{i_1=1}^{n_1}J_{\conf_0\to \conf_1^{i_1}} \right)\,.
\end{equation}
In this case $J_r$ is proportional to the entropy and thus exhibits a GCEM symmetry. However, we can also conclude in opposite direction that for systems obeying the constraint equations (\ref{consym}) in such a way that (\ref{Fkconst}) is not satisfied for all ${k}$, then the current $J_r$ is not proportional to $J_s$. In this case we expect to have a new type of symmetry different from the GCEM symmetry.

We point out that there is also a multi-dimensional version of the fluctuation theorem \cite{lebowitz99,andrieux07,faggionato11}, where the joint probability distribution of a set of currents (which when summed give the entropy) displays a symmetric large deviation function. It is
important to note that our symmetry is also different from this multi-dimensional case. This becomes clear if the fluctuation theorem obtained in \cite{andrieux07} using the cycle decomposition approach \cite{schnakenberg76} is considered. Following \cite{andrieux07}, the fact that the product of the
transition rates associated with a fundamental cycle in a given direction divided by the product of the transition rates in the opposite direction is
independent of the states within the cycle, is analogous to our condition (\ref{Fkconst}) holding for all columns.

\headline{Transitions network with odd $Q$}
So far we have considered transition networks shown in Fig. \ref{networkgeneral} with an even number of columns $Q$. The above results can be easily generalized to the case where $Q$ is odd. In this case there are at least two adjacent columns with only one configuration. 

In order to do this generalization we can consider an index $\tilde{k}$ that runs only over columns with more than one state. Note that we might also have odd columns with one state. Moreover we take $\tilde{Q}$ as the number of columns with more than one state, such that $\tilde{k}=1,\ldots,\tilde{Q}$. For the case where we have an even number of columns and all odd columns have more than one state we have $\tilde{Q}=Q/2$ and $k=2\tilde{k}-1$.

Note that the number of possible complete forward (and backward) cycles is given by $\prod_{\tilde{k}=1}^{\tilde{Q}}n_{\tilde{k}}$ and only the columns with more than one state are relevant in determining a complete cycle. Now if we consider the multiindex ${\multiindex i}=\{i_{\tilde{k}}\}$ and change definition (\ref{lambdabig}) to
\begin{equation}
\Lambda_{\multiindex i}\;=\;\Bigl\{\,\lambda(c_{\tilde{k}}^{j_{\tilde{k}}})\quad \Bigl|\quad \tilde{k}=1,2\ldots,\tilde{Q}; \quad j_{\tilde{k}}=1,\ldots,n_{\tilde{k}}; \quad j_{\tilde{k}}\neq i_{\tilde{k}}\Bigr\}
\end{equation}
the formulas (\ref{eqsymmetricfactor}), (\ref{defR}) and (\ref{consym}) take the same form if $Q$ is odd. Therefore, the proof also works for the $Q$ odd case.

\section{Examples}

In order to illustrate how this new symmetry can be established we consider some examples of physical systems described by jump processes with a network of states of the type shown in Fig. \ref{networkgeneral}, where the current $J_r$ has a clear physical meaning. The examples include a molecular motor in a particular network of states, the so-called restricted solid-on-solid (RSOS) model with four sites, and a growth model where the nucleation of the first particle and the completion of the layer take place in the time-scales of the same order.

\headline{Molecular motor with four different states}
A molecular motor is a biological protein that converts chemical energy into mechanical work by hydrolysis of adenosinetriphosphate (ATP) to adenosinediphosphate (ADP) and phosphate (P)~\cite{howard01} (see also \cite{lacoste} for a review in fluctuation relation for molecular motors).  Unlike macroscopic motors, which move unidirectionally in a well-defined cycle, a molecular motor performs a random walk driven by the chemical potential difference $\Delta \mu = \mu_{\rm ATP}-\mu_{\rm ADP}-\mu_{\rm P}$, moving preferentially forward if ATP is in excess. This allows one to model molecular motors by stochastic jump processes~\cite{juelicher97}. Assuming that the motor is thermally equilibrated with its local environment the jump rate is proportional to $\exp(\beta(\Delta \mu-W))$, where $W=Fl$ is the mechanical work to transport the cargo with force $F$ over the distance $l$.

\begin{figure}
\begin{center}	
\includegraphics[width=90mm]{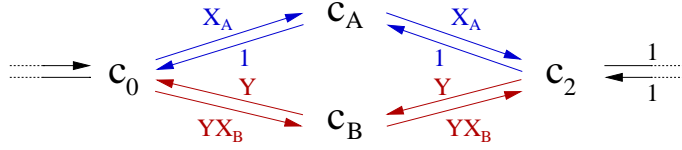}
\caption{Configuration network of a molecular motor with two possible transition cycles $A$ and $B$ (see text).} 
\label{network4state}
\end{center}
\end{figure}

The simplest model of a molecular motor would require a cycle of two states. In such a model the consumption of ATP would be proportional to the mechanical work. However, realistic molecular motors are often characterized by several possible transition cycles, i.e. the motor protein can advance by one step through different paths of intermediate configurations with different energy consumption. In what follows we consider a hypothetical network of four configurations with two transition paths denoted by $A$ and $B$, as shown in Fig. \ref{network4state}. One interesting question is whether such nano-machine leads to a better efficiency at maximum power than a simple linear chain \cite{seifert08}. Here we are interested in providing a simple example where our theory can be applied. Clearly, this network belongs to the class of networks studied in the previous section with $Q=3$ columns.

If both paths are characterized by different chemical potential differences $\Delta \mu_A$ and $\Delta \mu_B$, the jump rates in positive direction will be proportional to $X_{A}=\exp\left[\beta\left(\Delta\mu_{A}-Fl\right)\right]$ and $X_{B}=\exp\left[\beta\left(\Delta\mu_{B}-Fl\right)\right]$, respectively, while the rates for jump in opposite direction do not depend on the chemical potential. Moreover, we included a constant factor $Y$ in path $B$, accounting for a possibly different attempt frequency. Finally we assume that the transition rates between $\conf_0$ and $\conf_2$ are symmetric and equal to 1, as shown in Fig.~\ref{network4state}. The corresponding time evolution operator in a canonical basis $\{\conf_0,\conf_A,\conf_B,\conf_2\}$ reads
\begin{eqnarray}
\mathcal{L}=
\left(
\begin{array}{cccc}
 (1+X_{A}+YX_{B})		& -1		         & -Y	                  & -1 	\\
 -X_{A}		                     &(1+X_{A})	  &0	                  & -1	\\
-YX_{B}	                            &0	                &Y(1+X_{B})	&-Y	\\
-1	                                   &-X_{A}	       	  &-YX_{B}	        &2+Y	\\
\end{array}\right)\,.
\end{eqnarray}
We are now going to show that the propagation of the molecular motor defines a time-integrated current with a clear physical meaning that exhibits a symmetry in the large deviation regime that differs from the one for the entropy. To this end, we note that there are two possible complete cycles one going trough $\conf_A$ and the other trough $\conf_B$. The product of the rates along these cycles in forward and backward direction are given by
\begin{eqnarray}
T_A &=&\rate_{\conf_0\to \conf_A}\rate_{\conf_A\to \conf_2}\rate_{\conf_2\to \conf_0} \;=\; X_A^2\nonumber\\
\overline T_A &=& \rate_{\conf_0\to \conf_2}\rate_{\conf_2\to \conf_A}\rate_{\conf_A\to \conf_0} \;=\; 1\nonumber\\
T_B &=& \rate_{\conf_0\to \conf_B}\rate_{\conf_B\to \conf_2}\rate_{\conf_2\to \conf_0} \;=\; Y^2X_B^2\nonumber\\
\overline T_B &=& \rate_{\conf_0\to \conf_2}\rate_{\conf_2\to \conf_B}\rate_{\conf_B\to \conf_0} \;=\;Y^2
\end{eqnarray}       
Following the procedure described in the previous section we consider the time-integrated current (\ref{ourcurrent}). Note that the polynomials (\ref{fandfbar}) are of order one for the present model.  Comparing the leading order terms in Eq.~(\ref{constraints}) we obtain the condition
\begin{equation}
\exp(E \Theta)=\frac{T_A+T_B}{\overline{T}_A+\overline{T}_B}\,.
\end{equation}
Comparing the terms of order $x^0$ we obtain a constraint on the transition rates of the form
\begin{equation}
\frac{T_A+T_B}{\overline{T}_A+\overline{T}_B}=\frac{T_A\lambda_B+T_B\lambda_A}{\overline{T}_A\lambda_B+\overline{T}_B\lambda_A}\,,
\label{con4}
\end{equation} 
where $\lambda_A=X_A+1$ and $\lambda_B=Y(X_B+1)$ are the escape rate in the configurations $C_A$ and $C_B$, respectively. This equation has two solutions, namely $T_A/\overline{T}_A=T_B/\overline{T}_B$, for which (\ref{Fkconst}) holds so that the current is proportional to entropy in the large deviation regime, and $\lambda_A=\lambda_B$, which is the one that gives symmetry different from GCEM. Therefore, we need $\lambda_A=\lambda_B$, which implies 
\begin{equation}
Y=(1+X_A)/(1+X_B). 
\end{equation}
Without loss of generality we choose $\Delta\mu_A>\Delta\mu_B$, which gives $Y>1$. In this case, the above restriction on $Y$ means that the cycles that go trough configurations with higher chemical potential evolve at a slower time-scale. 

Assuming this condition to hold, we consider a time-integrated current of the form (\ref{ourcurrent}), namely the mechanical work
\begin{equation}
J_m \;=\; Fl\big(J_{\conf_A\to \conf_2}+J_{\conf_{0}\to \conf_A}+J_{\conf_B\to \conf_2}+J_{\conf_{0}\to \conf_B}\big)
\end{equation}
for which the factor~(\ref{eqsymmetricfactor}) is given by
\begin{equation}
E=\frac{1}{2Fl}\ln\frac{T_A+T_B}{\overline{T}_A+\overline{T}_B}=\frac{1}{2Fl}\ln\left[\frac{\left(X_{A}^{2}+X_{B}^{2}\right)\left(1+X_{B}\right)^{2}}{\left(1+X_{A}\right)^{2}+\left(1+X_{B}\right)^{2}}\right]\,.
\label{eqE}
\end{equation}
In order to demonstrate that this quantity exhibits a non-GCEM symmetry, we compute the smallest eigenvalues $\hat{I}_s(z)$ and $\hat{I}_m(z)$ of the modified time evolution operators (\ref{modgenerator}) for both the entropy
\begin{figure}
\begin{center}	
\includegraphics[width=85mm]{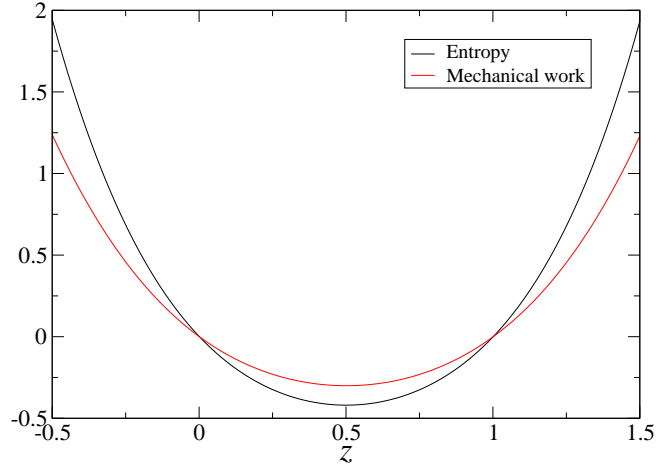}
\caption{Scaled cumulant generating functions for $F=\beta=l=1$, $\Delta\mu_A=3$, and $\Delta\mu_B=2$. The black line corresponds to the entropy and the red line corresponds to mechanical work  (with $z\to z/E$). They are obtained from the minimum eigenvalue of (\ref{entropymolecular}) and (\ref{mechmolecular}), respectively. They are both symmetric and not proportional.} 
\label{legendremechanical}
\end{center}
\end{figure}
\begin{eqnarray}
\hat{\mathcal{L}}_s(z)=
\left(
\begin{array}{cccc}
 (1+X_{A}+YX_{B})		& -X_A^{z} 		         & -YX_B^{z}	                  & -1 	\\
 -X_{A}^{(1-z)}		              &(1+X_{A})	  		  & 0	                  		    & -X_A^{z}	\\
-YX_{B}^{(1-z)}	                     &0	                		  &Y(1+X_{B})	           &-YX_B^{z}	\\
-1	                                   &-X_{A}^{(1-z)}	       	  	  &-YX_{B}^{(1-z)}	        &2+Y	\\
\end{array}\right)
\label{entropymolecular}
\end{eqnarray}
and for the mechanical work
\begin{eqnarray}
\hat{\mathcal{L}}_m(z)=
\left(
\begin{array}{cccc}
 (1+X_{A}+YX_{B})		& -1\exp(Flz) 		         & -Y\exp(Flz)	                  & -1 	\\
- X_{A}\exp(-Flz)		       &(1+X_{A})	  		  & 0	                  		    & -1\exp(Flz)	\\
-YX_{B}\exp(-Flz)	              &0	                		  &Y(1+X_{B})	           &-Y\exp(Flz)	\\
-1	                                   &-X_{A}\exp(-Flz)    	  	  &-YX_{B}\exp(-Flz)	           &2+Y	\\
\end{array}\right)\,.
\label{mechmolecular}
\end{eqnarray}
As expected, the large deviation function for the entropy obeys the GCEM symmetry $\hat{I}_s(z)=\hat{I}_s(1-z)$ while the mechanical work fulfills the symmetry $\hat{I}_m(z)=\hat{I}_m(E-z)$ with $E$ given in (\ref{eqE}). Moreover, plotting $I_s(z)$ with  $I_m(Ez)$ in Fig.~\ref{legendremechanical} we see that the rescaled Legendre transforms of the large deviation functions are different. Therefore, the mechanical work is a time-integrated current with a clear physical meaning that exhibits a new type of symmetry that differs from the GCEM symmetry of the entropy.

\begin{figure}[t]
\centering\includegraphics[width=155mm]{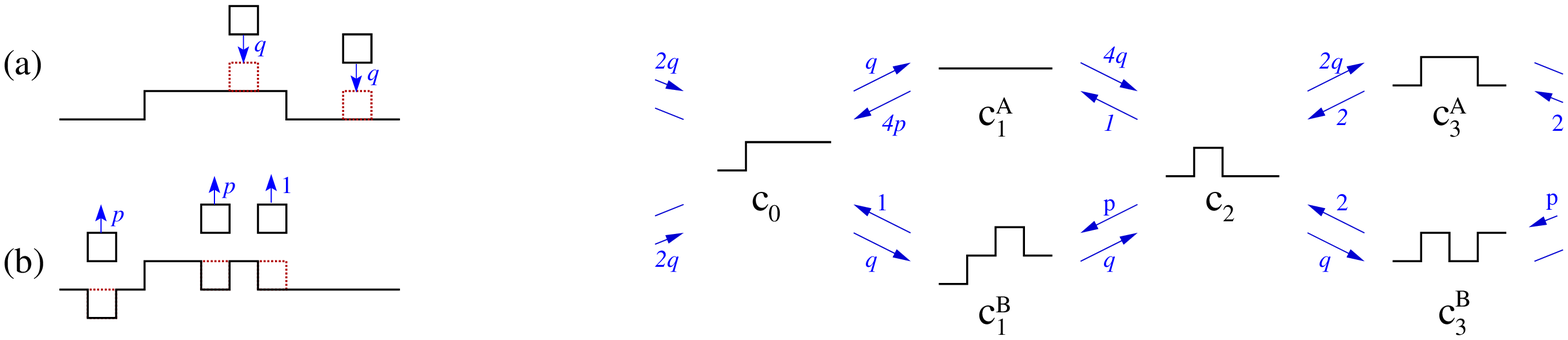}
\caption{Restricted solid on solid model. Left: Dynamic rules for (a) deposition and (b) evaporation. Right: Transition network of the model with four sites and periodic boundary conditions (see text).}
\label{rsos}
\end{figure}

\headline{Restricted solid on solid growth model in a four sites lattice}
The second example is a restricted solid-on-solid (RSOS) model for interface growth~\cite{barato10}. In this model the interface configuration is described by height variables $h_i\in\mathbb{Z}$ residing on the sites $i$ of a one-dimensional lattice. Particles are deposited everywhere with rate $q$ while they evaporate with rate~$1$ from the edges and with rate $p$ from the interior of plateaus, provided that the restriction $h_i-h_{i \pm 1}$ is not violated (see left panel of Fig.~\ref{rsos}). On an infinite lattice one observes that the interface roughens according to the predictions of the Kardar-Parisi-Zhang universality class \cite{KPZ}. 

Let us now consider a small system with $L=4$ sites with periodic boundary conditions. Moreover, let us identify all configurations which differ only by translation either in space or in height direction so that the interface can be in six possible configurations. By doing that, it can be demonstrated that the network of states takes the from shown in the right panel  Fig.~\ref{rsos} (see \cite{barato10} for details). This network belongs to the class of networks studied in the previous section. Moreover, one can easily show that the escape rates $\lambda(\conf_B^3)=\lambda(\conf_A^3)=2q+2$ in column $3$ are constant and that the rates across column $1$ obey the condition (\ref{Fkconst}),
\begin{equation}
\frac{\rate_{\conf_0\to \conf_1^A}\rate_{\conf_1^A\to \conf_2}}{\rate_{\conf_2\to \conf_1^A}\rate_{\conf_1^A\to \conf_0}}=\frac{\rate_{\conf_0\to \conf_1^B}\rate_{\conf_1^B\to \conf_2}}{\rate_{\conf_2\to \conf_1^B}\rate_{\conf_1^B\to \conf_0}}=\frac{q^2}{p}\,.
\end{equation}       
Hence, as explained in the previous section, the process fulfills the constraint (\ref{consym}). 

\begin{figure}[t]
\centering\includegraphics[width=80mm]{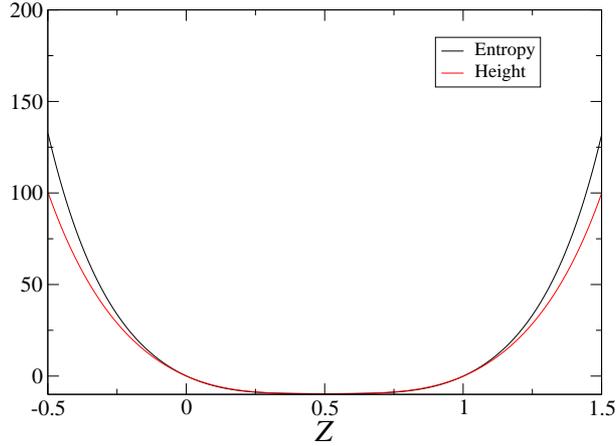}
\caption{Scaled cumulant generating functions for the RSOS model for $p=0.02$ and $q=10$. The red line corresponds to height  (with $z\to z/E$) and the black line to the entropy, they are obtained from the minimum eigenvalue of (\ref{matrixheightRSOS}) and (\ref{matrixentropyRSOS}), respectively.}
\label{legendreRSOS}
\end{figure}

The most important time-integrated current of the form (\ref{ourcurrent}) with a clear physical meaning is the total interface height which increases (decreases) by one whenever a deposition (evaporation) happens. The corresponding modified generator is given by
\begin{equation}
\hat{\mathcal{L}}_h(z)=
\left(\begin{array}{cccccc}
2q+p+2   & -4pe^{z} 	& -e^{z} 	& 0 	  	  & -2qe^{-z}	  	& -2qe^{-z} \\
-qe^{-z} & 4p+4q 	& 0 		& -e^{z} 	  & 0 	  		& 0 \\
-qe^{-z} & 0 	 	& q+1 		& -pe^{z} 	  & 0 	  		& 0 \\
0 	 & -4qe^{-z} 	& -qe^{-z} 	& 1+p+3q 	  & -2e^{z} 	  	& -2e^{z}\\
-2e^{z}  & 0 	 	& 0 		& -2qe^{-z} 	  & +2+2q 		& 0 \\
-pe^{z}  & 0 	 	& 0 		& -qe^{-z} 	  & 0 	  		& 2+2q
\end{array}\right)
\label{matrixheightRSOS}
\end{equation}
with the factor $E=\frac{1}{4}\ln\frac{3q^4}{2p+p^2}$, given by relation (\ref{eqsymmetricfactor}). Again its lowest eigenvalue $\hat I_h(z)$ has to be compared with the lowest eigenvalue $\hat I_s(z)$ of the corresponding modified generator for the entropy current, which reads
\begin{equation}
\hat{\mathcal{L}}_s(z)=
\left(\begin{array}{cccccc}
2q+p+2     		\,&\, -q^z(4p)^{1-z} 	\,&\, -q^z 		\,&\, 0 	  \,&\, -2q^{1-z}	  \,&\, -(2q)^{1-z}p^z \\
-q^{1-z}(4p)^{z}		\,&\, 4p+4q 	\,&\, 0 		\,&\, -(4q)^z 	  \,&\, 0 		  \,&\, 0 \\
-q^{1-z} 		\,&\, 0 	 	\,&\, q+1 		\,&\, -q^zp^{1-z}   \,&\, 0 		  \,&\, 0 \\
0 	  		\,&\, -(4q)^{1-z} 	 \,&\, -q^{1-z}p^z 	\,&\, 1+p+3q 	  \,&\, -2q^{z} 		  \,&\, -2^{1-z}q^z\\
-2q^{z}  		\,&\, 0 	 	\,&\, 0 		\,&\, -2q^{1-z}	  \,&\, 2+2q 		\,&\, 0 \\
-(2q)^zp^{1-z}  		\,&\, 0 	\,&\, 0 		\,&\, -2^zq^{1-z}   \,&\, 0 		  \,&\, 2+2q
\end{array}\right)
\label{matrixentropyRSOS}\,.
\end{equation}
Plotting the scaled cumulant generating function in Fig. \ref{legendreRSOS} we can see that both are symmetric and not proportional.
Unfortunately, for $L>4$ sites the height current is no longer symmetric because in this case the network of transitions is not of the form shown in Fig.~\ref{networkgeneral}. In the next example, we introduce a model where the space of states is of the form displayed in Fig. \ref{networkgeneral} for any system size.

\begin{figure}[t]
\centering\includegraphics[width=90mm]{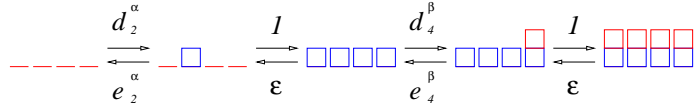}
\caption{Possible cycle of transitions for the case $L=4$. The $\alpha$ particles are blue and the $\beta$ particles red. The left-most configuration is a flat interface with $\beta$ particles.   
}
\label{growthmodelrules}
\end{figure}

\headline{Growth process with instantaneous monolayer completion}
As a third example, let us define a growth process with deposition and evaporation which has the special property that after a nucleation the actual monolayer is completed on a very short time scale. The model is defined on a one-dimensional lattice with particles of species $\alpha$ and $\beta$, assuming that only $\alpha$ particles can be deposited on top of $\beta$ particles and vice versa (see Fig. \ref{growthmodelrules}). Once a particle is deposited with rate $d_i^{\alpha,\beta}$ at site $i$ on a flat surface, the subsequent event is either the completion of the layer or the evaporation of the particle with rate $e_i^{\alpha,\beta}$. This describes a limit in which a monolayer is completed almost instantaneously after the first deposition. The transition rates for the completion of a layer is $1$ and the transition rate for the reversed event, that is, the evaporation of $L-1$ particles of a monolayer, is $\epsilon$.

The network of configurations of this model is of the form shown in Fig. \ref{networkgeneral} with $Q=4$, where $\conf_0$ and $\conf_2$ correspond to flat interface configurations while the columns $\conf_1$ and $\conf_3$ have $L$ states, each corresponding to a single particle nucleated at one of the $L$ sites. In this model the physically relevant time-integrated current of the form (\ref{ourcurrent}) is the interface height, which increases by $1/L$ when a particle is deposited on a flat interface, and by $(L-1)/L$ whenever a monolayer is completed.

\begin{figure}[b]
\centering\includegraphics[width=80mm]{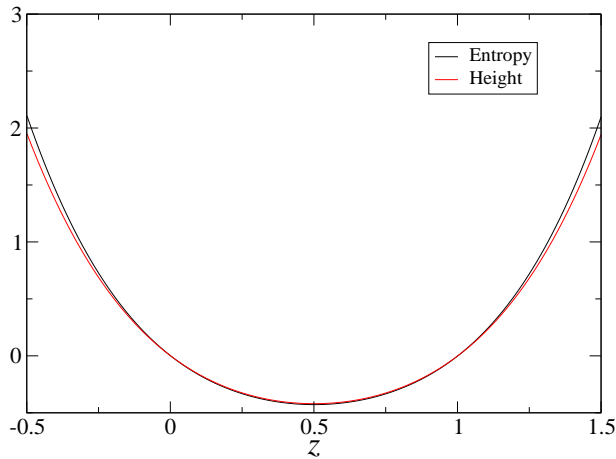}
\caption{Scaled cumulant generating functions for the growth model with fast monolayer completion. We used $L=3$, $d^\alpha_1=1$, $d^\alpha_2=2$, $d^\alpha_3=3$, $d^\beta_1=2$, $d^\beta_2=2$, $d^\beta_3=5$, $e^\alpha_1=e^\alpha_2=e^\alpha_3=2$, $e^\beta_1=e^\beta_2=e^\beta_3=3$, and $\epsilon=0.05$. The black line is related to entropy and the red line to height.
}
\label{legendreDavid}
\end{figure}

The product of rates on a forward and backward cycle through $\conf_1^i$ and $\conf_3^j$ are $T_{ij}= d^\alpha_id^\beta_j$ and $\overline{T}_{ij}= \epsilon^{2}e^\alpha_ie^\beta_j$ while the escape rates are $\lambda(\conf_1^i)=\lambda_i^\alpha= 1+e_i^\alpha$ and $\lambda(\conf_3^i)=\lambda_i^\beta= 1+e_i^\beta$. Hence, the polynomials (\ref{fandfbar}) with degree $2L-2$ are given by 
\begin{eqnarray}
f(x)&=&\sum_{i=1}^{L}\sum_{j=1}^{L}d_i^{\alpha}d_j^{\beta}\prod_{l\neq i}(x-e_l^{\alpha}-1)\prod_{m\neq j}(x-e_m^{\beta}-1)\\
\overline{f}(x)&=&\epsilon^2\sum_{i=1}^{L}\sum_{j=1}^{L}e_i^{\alpha}e_j^{\beta}\prod_{l\neq i}(x-e_l^{\alpha}-1)\prod_{m\neq j}(x-e_m^{\beta}-1)
\end{eqnarray}
and for the factor (\ref{eqsymmetricfactor}) we get
\begin{equation}
E= \ln\frac{\sum_{i,j=1}^{L}d_i^{\alpha}d_j^{\beta}}{\epsilon^{2}\sum_{i,j=1}^{L}e_i^{\alpha}e_j^{\beta}}.
\label{Eheight}
\end{equation}   
The constraints (\ref{consym}) take the form
\begin{equation}
\frac{\sum_{i,j=1}^{L}d_i^{\alpha}d_j^{\beta}}{\epsilon^{2}\sum_{i,j=1}^{L}e_i^{\alpha}e_j^{\beta}}=
\frac{\sum_{i,j=1}^{L}d_i^{\alpha}d_j^{\beta}\sigma_k\left[\lambda_1^\alpha,\ldots,\lambda_{i-1}^\alpha,\lambda_{i+1}^\alpha,\ldots,\lambda_L^\alpha,\lambda_1^\beta,\ldots,\lambda_{j-1}^\beta,\lambda_{j+1}^\beta,\ldots,\lambda_L^\beta\right]}{\epsilon^{2}\sum_{i,j=1}^{L}e_i^{\alpha}e_j^{\beta}\sigma_k\left[\lambda_1^\alpha,\ldots,\lambda_{i-1}^\alpha,\lambda_{i+1}^\alpha,\ldots,\lambda_L^\alpha,\lambda_1^\beta,\ldots,\lambda_{j-1}^\beta,\lambda_{j+1}^\beta,\ldots,\lambda_L^\beta\right]}
\label{constraintDavid},
\end{equation}   
where $k=1,2,\ldots,2L-2$. Whenever the deposition and evaporation rates fulfill the above equation, the probability distribution of velocity will symmetric with respect to (\ref{Eheight}). Again, this symmetry is generally different from the GCEM symmetry. As an example in Fig. \ref{legendreDavid} we plot the scaled cumulant generating function for the entropy and height for a system with $L=3$ sites, where the transition rates are chosen in such a way that the escape rate is constant in columns $\conf_1$ and $\conf_3$ so that the relation (\ref{constraintDavid}) is satisfied.

\section{On the origin of the symmetry}

In the following we demonstrate that the new symmetry we proved here also comes from time-reversal. However, in order to see the symmetry we have to consider a functional of a group of trajectories instead of a single trajectory. We will restrict our analysis to the four-state system shown in Fig.~\ref{network4state}, but we expect the same kind of proof to be valid for general networks of the form shown in Fig .~\ref{networkgeneral}. Before going into detail, we provide a simple demonstration of the GCEM symmetry, where it becomes clear that its physical origin is related to time-reversal.

\headline{A simple demonstration of the fluctuation theorem}
As in Sect. 2 we consider stochastic trajectory with $M$ jumps in the time interval $[t_0,t_f]$ and jumps from $\conf(t_{i})$ to $\conf(t_{i+1})$ take place at times $t_i$. We denote this trajectory by $\overrightarrow{C}_{M,t}$ and its weight is given by  
\begin{equation}
W[\overrightarrow{C}_{M,t}]= \exp[-\lambda(\conf(t_M))(t_f-t_M)]\prod_{i=0}^{M-1}\rate_{\conf(t_{i})\to \conf(t_{i+1})}\exp[-\lambda(\conf(t_i))(t_{i+1}-t_i)],
\end{equation}
where $\lambda(\conf)$ is the scape rate from state $\conf$. We note that we should also multiply the weight by the probability distribution of the initial state, but we are considering a uniform initial distribution of states. The reversed trajectory, where the system starts at state $\conf(t_M)$ at time $t_f$ (again with the uniform distribution) and jumps from state $\conf(t_{i+1})$ to state $\conf(t_{i})$ at time $t_f+t_0-t_{i+1}$, is denoted by $\overleftarrow{C}_{M,t_f+t_0-t}$. The weight of this trajectory is written as
\begin{equation}
W[\overleftarrow{C}_{M,t_f+t_0-t}]= \exp[-\lambda(\conf(t_M))(t_f-t_M)]\prod_{i=0}^{M-1}\rate_{\conf(t_{i+1})\to \conf(t_{i})}\exp[-\lambda(\conf(t_i))(t_{i+1}-t_i)].
\end{equation}
The entropy current (\ref{entropy}) is related to the weight of a trajectory divided by the weight of the time-reversed trajectory by
\begin{equation}
\exp(-J_s[\overrightarrow{C}_{M,t}]])= \frac{W[\overleftarrow{C}_{M,t_f+t_0-t}]}{W[\overrightarrow{C}_{M,t}]},
\label{entropypath}
\end{equation}
From (\ref{entropypath}) it follows that 
\begin{equation}
W[\overrightarrow{C}_{M,t}]\exp(-X)\delta(J_s[\overrightarrow{C}_{M,t}]-X)= W[\overleftarrow{C}_{M,t_f+t_0-t}]\delta(J_s[\overrightarrow{C}_{M,t}]-X)\,.
\end{equation}
Summing over all possible trajectories and using $J_s[\overrightarrow{C}_{M,t}]=-J_s[\overleftarrow{C}_{M,t_f+t_0-t}]$ we obtain the fluctuation theorem (\ref{GCEMforlarge}), i.e.,
\begin{equation}
\frac{P(J_s=-X)}{P(J_s=X)}=\exp(-X).
\end{equation}
This demonstrates that the GCEM symmetry is a direct consequence of the fact that the entropy is the weight of a trajectory divided by the weight of the time-reversed trajectory.

\headline{Time-reversal of a group of trajectories}
We now consider Markov jump process with the network of states given in Fig. \ref{network4state}. In the following we will denote the states $\conf_A$ and $\conf_B$ by $\conf_1^A$ and $\conf_1^B$, respectively. Here, instead of considering one stochastic trajectory, we consider a certain group of trajectories (or class of trajectories), which can be defined as follows. Two trajectories belong to the same class if they have the same number of jumps taking place at the same times $t_i$, following of the same sequence of columns. For example, the trajectories $\conf_0\to \conf_1^A\to \conf_2\to \conf_0$ and $\conf_0\to \conf_1^B\to \conf_2\to \conf_0$ belong to the same class. More generally, a trajectory that goes trough column $1$ for $K$ different times pertains to a class with $2^K$ trajectories. 

We denote this group of trajectories by $\{\overrightarrow{C}_{M,t}\}$ and its weight, which is the sum of the weights of each trajectory in the group, by $R[\{\overrightarrow{C}_{M,t}\}]$. We define a quantity $\tilde{J}$ such that
\begin{equation}
\exp(-{\tilde{J}}[\{\overrightarrow{C}_{M,t}\}])= \frac{R[\{\overleftarrow{C}_{M,t_f+t_0-t}\}]}{R[\{\overrightarrow{C}_{M,t}\}]},
\label{current2}
\end{equation}
where $R[\{\overleftarrow{C}_{M,t_f+t_0-t}\}]$ is the sum of the weights of the reversed trajectories in the group and are using a tilde to denote functionals of the group of trajectories. Note that, unlike the entropy, the current $\tilde{J}$ is a functional of a group of trajectories. What we show next is that in the case $\lambda_1=\lambda(\conf_1^A)=\lambda(\conf_1^B)$ (which is the sufficient condition (\ref{constressimple}) for the symmetry), then the the functional $\tilde{J}_r$, which is generated by the current $J_r$ in the way explained below, is equal to $\tilde{J}$, provided the increments $\theta_k$ defined in (\ref{ourcurrent}) are chosen appropriately.  

The difference between a current that is a functional of the stochastic path $\overrightarrow{C}_{M,t}$ and a current that is a functional of the group of paths $\{\overrightarrow{C}_{M,t}\}$ is in the increment when there is a jump to an state at column $1$. For the second kind of currents the increment is as follows. Let us consider a trajectory which at time $t_{i-1}$ hops to a state $c(t_i)$ that pertains to the column $c_1$, and stays in that state during the time interval $t_i - t_{i-1}$. The contribution to the current $\tilde{J}[\{\overrightarrow{C}_{M,t}\}]$ is given by
\begin{equation}
\ln\frac{\rate_{\conf(t_{i-1})\to \conf_1^A}\rate_{\conf_1^A\to \conf(t_{i+1})}\exp[-\lambda(\conf_1^A)(t_{i}-t_{i-1})]+\rate_{\conf(t_{i-1})\to \conf_1^B}\rate_{\conf_1^B\to \conf(t_{i+1})}\exp[-\lambda(\conf_1^B)(t_{i}-t_{i-1})]}{\rate_{\conf(t_{i+1})\to \conf_1^A}\rate_{\conf_1^A\to \conf(t_{i-1})}\exp[-\lambda(\conf_1^A)(t_{i}-t_{i-1})]+\rate_{\conf(t_{i+1})\to \conf_1^B}\rate_{\conf_1^B\to \conf(t_{i-1})}\exp[-\lambda(\conf_1^B)(t_i-t_{i-1})]}.
\end{equation}
If the escape rate is constant at column $1$ the above term becomes
\begin{equation}
\ln\frac{\rate_{\conf(t_{i-1})\to \conf_1^A}\rate_{\conf_1^A\to \conf(t_{i+1})}+\rate_{\conf(t_{i-1})\to \conf_1^B}\rate_{\conf_1^B\to \conf(t_{i+1})}}{\rate_{\conf(t_{i+1})\to \conf_1^A}\rate_{\conf_1^A\to \conf(t_{i-1})}+\rate_{\conf(t_{i+1})\to \conf_1^B}\rate_{\conf_1^B\to \conf(t_{i-1})}}.
\end{equation}
Note that, the same relation holds if the relation (\ref{Fkconst}) is valid at column $1$. The fact that the scape rates are constant simplify the situation considerably: in this case the dependence on the time interval $t_i-t_{i-1}$ disappears. From now on we assume that the escape rates are constant at column $1$.

The current $J_r$ is a functional that is invariant within the class of trajectories, i.e., all the trajectories in the same group have the same value of $J_r$. This important property comes from the fact the the current  (\ref{ourcurrent}) is defined in a way such that it does not discriminate between different states in the same column. Hence, $J_r$ induces a current $\tilde{J}_r$ defined on a class of trajectories by the relation  
\begin{equation}
\tilde{J}_r[\{\overrightarrow{C}_{M,t}\}]=J_r[ \overrightarrow{C}_{M,t}] \,\,\mbox{for}\,\, \overrightarrow{C}_{M,t}\in\{\overrightarrow{C}_{M,t}\}.
\label{defJt}
\end{equation} 
If we consider the current $J_r$ with the increments   
\begin{equation}
\theta_0=\theta_1=\frac{1}{2}\ln\frac{\rate_{\conf_0\to \conf_1^A}\rate_{\conf_1^A\to \conf_2}+\rate_{\conf_0\to \conf_1^B}\rate_{\conf_1^B\to \conf_2}}{\rate_{\conf_2\to \conf_1^A}\rate_{\conf_1^A\to \conf_0}+\rate_{\conf_2\to \conf_1^B}\rate_{\conf_1^B\to \conf_0}}\,,\qquad
\theta_2= \ln\frac{\rate_{\conf_2\to \conf_0}}{\rate_{\conf_0\to \conf_2}},
\label{incre}
\end{equation} 
then it is clear that $\tilde{J}_r[\{\overrightarrow{C}_{M,t}\}]$ is equal to the  current given in (\ref{current2}). It then follows that 
\begin{equation}
R[\{\overrightarrow{C}_{M,t}\}]\exp(-X)\delta(\tilde{J}_r[\{\overrightarrow{C}_{M,t}\}]-X)= R[\{\overleftarrow{C}_{M,t_f+t_0-t}\}]\delta(\tilde{J}_r[\{\overrightarrow{C}_{M,t}\}]-X)\,.
\end{equation}
Now summing over all possible group of trajectories we get $\frac{P(\tilde{J}_r=-X)}{P(\tilde{J}_r=X)}=\exp(-X)$. Since relation (\ref{defJt}) gives 
$P(\tilde{J}_r=X)=P(J_r=X)$ we finally obtain 
\begin{equation}
\frac{P(J_r=-X)}{P(J_r=X)}=\exp(-X).
\end{equation}
This last relation implies the symmetry for the current $J_r$ with the choice of increments (\ref{incre}). However, as we proved in (\ref{1cld}), at large time all the currents $J_r$ are proportional for any choice of increments $\theta_k$. Therefore, this argument shows that the symmetry in the current  $J_r$ comes also from time-reversal, but at a more coarse grained level. We point out that functionals of a group of paths have been considered also in \cite{rahav}, however in a rather different context.  

We believe that the same demonstration can be extended for the more  general network of Fig. \ref{networkgeneral}. However, formalizing the proof for this more general case leads to much more cumbersome formulas. We pretend to address this problem in future work.  

\section{Conclusion}

In this paper we have investigated a new symmetry of time-integrated currents which is generally different from the GCEM-symmetry of the entropy current in the long time limit. It is valid for a very restricted class of Markov processes in the sense that they have the very peculiar network of configurations shown in Fig. \ref{networkgeneral}. Moreover, the symmetry appears only when the transition rates fulfill a certain set of constraints. Nevertheless, to our knowledge, this is the only case where a current not proportional to the entropy in the large deviation regime displays a symmetric large deviation function.
 
We showed three physical examples where this current is a relevant physical observable. We considered a toy model for a molecular model, where the mechanical work has a symmetric large deviation function different from the one for the entropy. Moreover, we analyzed two growth models where the height displays such a symmetry.

As is the case of the GCEM symmetry, the origin of our symmetry seems to be related to time-reversal. However, as we showed in Sec 5 for a $4$ states system, the symmetry becomes clear only when we consider a functional of a group of stochastic trajectories. That is, the symmetric current as a functional of the group of trajectories is given by the weight of the trajectories divided by the weight of the reversed trajectories. We demonstrated the symmetry with this grouping of trajectories argument only for the specific $4$ states system, but we expect the same kind of proof also to be valid for the whole class of networks considered here.  

A natural extension of the this work is to look for other currents, in a more general space of states,  with a symmetric large deviation function. This can be done by looking for the conditions that the transition rates have to fulfill in order for the characteristic polynomial of the modified generator to be symmetric. We expect that whenever this characteristic polynomial is symmetric, the origin of the symmetry should be related to time-reversal (of a trajectory or a group of trajectories). One might speculate about a situation where the characteristic polynomial of the modified generator is not symmetric, however its minimum eigenvalue is still symmetric. In this case the origin of the symmetry might be also related to time-reversal, but of some most probable stochastic trajectory dominating a sum over different trajectories. Ultimately, it would be of great theoretical interest for nonequilibrium statistical physics to find out which are the  time-integrated currents with a symmetric large deviation function and what might be the origin of these symmetries.

\begin{acknowledgements}
The support of the Israel Science Foundation (ISF) is gratefully acknowledged. ACB and RC thank the Weizmann Institute of Science for hospitality. 
\end{acknowledgements}



\begin{thebibliography}{10}


\bibitem{evans93}  Evans, D. J., Cohen, E. G. D., Morriss G. P.:  Phys. Rev. Lett. \textbf{71}, 2401 (1993)

\bibitem{evans94}  Evans, D. J., Searles, D. J.:  Phys. Rev. E \textbf{50}, 1645 (1994)

\bibitem{gallavotti95}  Gallavotti, G., Cohen, E. G. D.: Phys. Rev. Lett. \textbf{74}, 2694 (1995)

\bibitem{kurchan98}  Kurchan, J.:  J. Phys. A: Math. Gen. \textbf{31}, 3719 (1998)

\bibitem{lebowitz99}  Lebowitz, J. L., Spohn, H.: J. Stat. Phys. \textbf{95}, 333 (1999)

\bibitem{maes99}  Maes, C.: J. Stat. Phys. \textbf{95}, 367 (1999)

\bibitem{Jia1}  Jiang, D-Q., Qian, M., Qian, M-P.: Mathematical Theory of Nonequilibrium Steady States. Berlin, Springer (2004)

\bibitem{andrieux07}  Andrieux, D., Gaspard, P.: J. Stat. Phys. \textbf{127}, 107 (2007)

\bibitem{harris07}  Harris, R. J., Sch\"{u}tz, G. M.: J. Stat. Mech., P07020 (2007)

\bibitem{kurchan07}  Kurchan, J.: J. Stat. Mech., P07005 (2007)

\bibitem{zia07}  Zia, R., Schmittmann, B.: J. Stat. Mech., P07012 (2007)

\bibitem{schnakenberg76}  Schnakenberg, J.:  Rev. Mod. Phys. \textbf{48}, 571 (1976)

\bibitem{seifert05}  Seifert, U.:  Phys. Rev. Lett. \textbf{95}, 040602 (2005)

\bibitem{jarzynski97}  Jarzynski, C.: Phys. Rev. Lett. \textbf{78}, 2690 (1997)

\bibitem{croocks99}  Crooks, G. E.: Phys. Rev. E \textbf{60}, 2721 (1999)

\bibitem{ellis85}  Ellis, R. S.: Entropy, Large Deviations, and Statistical Mechanics. New York, Springer (1985)

\bibitem{hollander}  den Hollander, F.: Large Deviations. Providence, Field Institute Monographs (2000)

\bibitem{touchette09}  Touchette, H.: Phys. Rep. \textbf{478}, 1-69 (2009)

\bibitem{oono} Oono, Y.:  Progr. Theoret. Phys. Suppl. \textbf{99}, 165-205 (1989)

\bibitem{barato10} Barato, A.C., Chetrite, R., Hinrichsen, H., Mukamel, D.: J. Stat. Mech., P10008 (2010)

\bibitem{hinrichsen11} Hinrichsen, H., Gogolin, C., and Janotta, P., J. Phys.: Conf. Ser. \textbf{297}, 012011 (2011)

\bibitem{bodineau07}  Bodineau, T., Derrida, B.: C. R. Physique \textbf{8}, 540-555 (2007)

\bibitem{faggionato11} Faggionato A., Di Pietro D.: J. Stat. Phys. \textbf{143}, 11-32 (2011) 

\bibitem{howard01} Howard, J.: Mechanics of Motor Proteins and the Cytoskeleton. 1st edition Sinauer, New York (2001)

\bibitem{lacoste} Lacoste, D., Mallick, K.: Fluctuation relations for molecular motors. In: Duplantier, B., Rivasseau, V. (eds.) Biological Physics. Poincar\'e Seminar 2009. Progress in Mathematical Physics, vol. 60. Birkhäuser, Basel (2011) 

\bibitem{juelicher97} J{\"u}licher, F., Rev. Mod. Phys. 69, 1269 (1997) 

\bibitem{seifert08} Schmiedl, T., Seifert, U.: Euro Phys. Lett., \textbf{83},  (2008)

\bibitem{KPZ}  Kardar, M., Parisi, G., Zhang, Y. -C.: Phys. Rev. Lett. \textbf{56}, 889 (1986)

\bibitem{rahav}  Rahav, S., Jarzynski, C.: J. Stat. Mech., P09012 (2007)


\end{thebibliography}
\end{document}